\documentstyle[aps,manuscript,psfig]{revtex}

\begin{document}

\titlepage
\setcounter{page}{1}
\title{
Hyperon polarization and single spin left-right asymmetry 
in inclusive production processes at high energies}
\author {{Liang Zuo-tang$^1$, and C. Boros$^{2,3}$}}
\address {{$^1$Department of physics, Shandong University, Jinan, Shandong 
250100, China\\
$^2$Institut f\"ur Theoretische Physik,
Freie Universit\"at Berlin,
Arnimallee 14, 14195 Berlin, Germany\\
$^3$Special Research Centre for the Subatomic Structure of Matter,
University of Adelaide, Adelaide, Australia 5005
}}
\author{(ADP-97-41/T269 \,\,\,\,  hep-ph/9708488)}

\maketitle

\begin{abstract}                
It is shown that the polarization of hyperons 
observed in high energy collisions using  
unpolarized hadron beams and unpolarized 
nucleon or nuclear targets 
is closely related to the left-right asymmetries 
observed in single spin inclusive hadron production processes.
The relationship is most obvious for the production 
of the hyperons which have only one common 
valence quark with the projectile. 
Examples of this kind are given. 
Further implications of the
existence of large polarization for hyperon 
which has two valence quarks in common with the projectile 
and their consequences are discussed.
A comparison with the available data is made. 
Further tests are suggested.
\end{abstract}     


\newpage

Since the discovery of the 
striking hyperon polarization ($P_H$) in 
inclusive production processes at high energies [1], 
there has been much interest in studying 
the origin of this effect, 
both experimentally [2] and theoretically [3].  
It is now a well-known fact [1-2] that hyperons produced 
in high energy hadron-hadron collisions
are polarized transversely to the production plane,  
although neither the projectiles nor  
the targets are polarized. 
Experimental results for  
different kinds of hyperons 
in different reactions at different energies show 
the following striking characteristics: 
(1) $P_H$ is significant in, 
    and only in, the fragmentation regions of the colliding objects;
(2) $P_H$ depends on the flavor quantum numbers of the produced hyperon;
(3) $P_H$ in the projectile fragmentation 
    region depends very much on 
    the flavor quantum numbers of the projectile but  
    little on those of the target.

More recently, 
striking left-right asymmetries ($A_N$) 
have been observed [4-7]  
in single-spin hadron-hadron collisions.  
The available data 
for inclusive production of 
different mesons and of $\Lambda$ hyperon 
show very much the 
same characteristics as those for $P_H$: 
we can simply replace $P_H$ by $A_N$ 
in (1)-(3) above. 
Not only these striking similarities but also 
the following reasonings strongly suggest 
that these two phenomena 
should be closely related to each other.
We note: $A_N\not=0$ implies that  
the direction of transverse 
motion of the produced hadron depends on  
the polarization of the projectile. 
E.g., for $\pi^+$ in 
$p(\uparrow)+p\to \pi^++X$, $A_N>0$ [5,6];  
and this means that the produced $\pi^+$ has large 
probability to go left looking down stream if the 
projectile is upwards polarized.   
$P_H\not=0$ means that there exists a 
correlation between the direction of transverse 
motion of the produced hyperon ($H$) and 
the polarization of this hyperon. 
We recall that $P_H$ is 
defined with respect to (w.r.t.) the production plane 
and, e.g., $P_\Lambda<0$ in $p+p\to \Lambda+X$ means that 
the $\Lambda$'s which are going left (looking down stream) 
have a larger probability to be downwards polarized.  
We see, both phenomena show the existence of 
a correlation between transverse motion 
and transverse polarization.  
Hence, unless we insist on {\it assuming} that 
the polarization of 
the produced hyperons in 
the projectile fragmentation region  
is independent of that of the projectile 
--- which would in particular contradict the empirical  
fact recently  observed  by E704 
Collaboration [6] for $\Lambda$ production --- 
we are practically forced to accept that 
$A_N$ and $P_H$ are closely related to each other.

The close relation between $A_N$ 
and $P_H$ is most obvious  
in the case in which 
the produced hyperon ($H$) has only 
one valence quark in common with the projectile. 
In this case, the beam fragmentation region 
is dominated by the hadronization product that contains 
this common valence quark. 
To see whether such hyperon is polarized and, if yes, 
how large $P_H$ is, we recall the following:

(I) The existence of $A_N$ in single-spin reactions 
shows that the polarization of the valence quark and the 
transverse moving direction of the produced hadron 
 containing this valence quark
are closely related to each other:
the data [4-7] shows that 
meson (e.g. $\pi$, $\eta$ or $K$) 
 containing $q_v^P$ 
and a suitable anti-sea-quark $\bar q_s^T$ 
has a large probability to go left 
if $q_v^P$ is upwards polarized.
(Here, $P$ or $T$  
denotes projectile or target,
 $v$ or $s$  valence or sea).    
Hence, if the produced meson is going left, 
the corresponding $q_v^P$ should have a
large probability to be upwards polarized. 
We assume that this is also true for the produced baryon 
which contains such a $q_v^P$ and a sea diquark. 

(II) Recent measurement [8] 
of $\Lambda$ polarization from $Z^0$ decay 
by ALEPH Collaboration shows that,
in the longitudinally polarized case,  
quark polarization remains the same 
before and after the hadronization. 
We assume that this is true 
not only for $\Lambda$ production 
but also for other hyperons 
and also in the transversely polarized case. 

We note that both (I) and (II) are direct extensions of 
the experimental observations. 
They can also be directly tested  
by performing further experiments,  
e.g. by measuring $A_N$ 
in $p(\uparrow)+p\to H+X$ 
and $P_H$ in the current fragmentation region of 
$e^-+p(\uparrow)\to e^-+H+X$ 
for $H=\Sigma^-$ (or $\Xi^0$, or $\Xi^-$) 
respectively.  
[Here, $p(\uparrow)$ denotes a transversely polarized proton.]
Theoretically, 
whether (II) is true depends on the detailed mechanism 
of hadronization, which is in general of soft nature 
and at present can only be described using 
phenomenological models. 
It can easily be seen that (II) is indeed true 
in the popular models such as the LUND model [9]. 
The validity of (I) 
has been a puzzle for a long time 
and a number of models have been proposed [10] recently,  
which can give rise to such $A_N$'s. 
Yet, which one is more appropriate is still in debate. 
Since the purpose of this paper is to discuss the 
relationship between $P_H$ and $A_N$ independent 
of these models, 
we will just take (I) and (II) as assumptions and  
show that $P_H$ in unpolarized $pp$-collisions 
can be determined uniquely using these two points. 
This is quite straight forward: 
Since $P_H$ is defined w.r.t. the production plane, 
we need only to consider e.g. those hyperons 
which are going left and check 
whether they are upwards (or downwards) polarized.
According to (I), if the hyperon is going left, 
$q_v^P$ should have a large probability to be upwards polarized.
This means, by choosing those hyperons which are going left, 
we obtain a subsample of hyperons which are formed by $q_v^P$'s 
that are upwards polarized with suitable sea diquarks.  
According to (II), these valence quarks 
remain upwards polarized in 
the produced hyperons. 
This, together with the wave function of the hyperon, 
determines whether the hyperons are polarized and, 
if yes, how large the polarizations are. 
To demonstrate this explicitly, 
we consider $p+p\to \Sigma^-+X$. 
Here, the dominating contribution  
in the fragmentation region is the $\Sigma^-$ 
made out of the common valence quark $d_v^P$ 
and a sea diquark $(d_ss_s)^T$; and  
the wave function of $\Sigma^-$ is: 
$|\Sigma ^{-\uparrow} \rangle ={1\over 2\sqrt{3}}
 [3d^\uparrow (ds)_{0,0}+ d^\uparrow (ds)_{1,0}
  -\sqrt{2} d^\downarrow(ds)_{1,1} ]$,
where the subscripts of the diquarks are their total   
angular momenta and the third components.
We see that if $d_v^P$ is upwards polarized, 
$\Sigma^-$ has a probability of 
$5/6$ ($1/6$) to be upwards (downwards) polarized. 
Hence, we obtain that the $\Sigma^-$ which contains 
the $d_v^P$ and a $(d_ss_s)^T$ is 
positively polarized and the polarization 
is $(5/6)C$,  
(where $0<C<1$ is the difference [11,12] 
between the probability for $B$ made out of
$q^P_v$ and $(q_sq_s)^T$ to 
go left and that to go right if $q_v^P$ is upwards polarized.) 
Similar analysis can also be done for other hyperons.
We obtain e.g. that both $\Xi^-$ and $\Xi^0$ produced 
in $pp$-collisions are negatively polarized and 
the polarization is $-C/3$, which implies 
that their magnitudes are smaller 
than that of $P_{\Sigma^-}$. 
Since hyperons containing the $q_v^P$'s 
dominate only at large $x_F$
($x_F\equiv 2p_\parallel /\sqrt{s}$, where $p_\parallel$ is 
the longitudinal momentum of the produced hyperon, $\sqrt{s}$ 
is the total c. m. energy of the colliding hadron system.)
we expect that the magnitudes of $P_H$ 
increase with increasing $x_F$ and the 
above mentioned results are their 
limits at $x_F\to 1$. 
All these are consistent with the data [1,2].

Without any other input, we obtained 
also many further direct associations, in particular 
the following: 
(A) {\it $P_\Lambda$ in the beam fragmentation region 
    of $K^-+p\to \Lambda+X$ is large and is, in contrast to 
    that in $pp$-collisions, positive in sign.}
This is because, according to the wave function, 
$|\Lambda^\uparrow \rangle =s^\uparrow (ud)_{0,0}$, 
the polarization of $\Lambda$ is entirely determined 
by the $s$ quark.
Here, the dominating contribution 
is the $\Lambda$ which contains the $s_v^P$ of $K^-$ 
and a suitable $(u_sd_s)^T$;
and $s^P_v$ should have large probability to be upwards polarized 
if $\Lambda$ goes left.
(B) {\it $P_\Lambda$ in the beam fragmentation region of 
  $\pi^\pm+p\to \Lambda +X$ should be negative 
  and the magnitude should be very small.}
This is because, the dominating contribution here  
is the $\Lambda$ containing the $u^P_v$ (or $d_v^P$) 
of $\pi^+$ (or $\pi^-$) and 
a suitable $(d_ss_s)^T$ [or a $(u_ss_s)^T$]. 
Although the $u_v^P$ (or $d_v^P$) should 
have a large probability to be upwards polarized 
if $\Lambda$ goes left, 
$\Lambda$ itself remains unpolarized,  
since its polarization is determined 
solely by the $s$ quark.   
A small $P_\Lambda$ is expected 
only from the decay of $\Sigma^0$.
(C){\it Not only hyperons 
 but also the produced vector mesons 
 are expected to be transversely polarized
 in the fragmentation region of hadron-hadron collisions.}
E.g., $\rho ^\pm, \rho^0, K^{*+}$ 
in the fragmentation regions of $pp$-collisions 
are expected to be positively polarized. 
This is because, 
the dominating contribution here is the meson  
containing a $q^P_v$ and a $\bar q_s^T$, and  
the $q_v^P$ should have a large probability 
to be upwards polarized if the meson is going left. 
(D) {\it Neither the contribution from 
hadronization to $P_\Lambda$ nor 
that to $A_N$ can be large.}  
The former is a direct implication of the 
results of measurements [8,13] in 
$e^+e^-\to\Lambda+X$, 
which show no significant 
transverse polarization $P_\Lambda$.  
The close relation between $A_N$ and $P_H$ 
implies that the latter should also be true.
Presently, there are already data available 
for the processes mentioned 
in (A) and (B) [14,15], 
and both of them are {\it in agreement with 
these associations}. 
(D) is consistent with the results [16] of the recent 
SLD measurements of jet handedness at SLAC, which show 
that the spin dependence of hadronization is very little.
(C) can be checked by future experiments.

Encouraged by these agreements, 
we continue to discuss the second case in which the 
produced hyperon has two valence quarks 
in common with the projectile and hence
hyperons containing such common valence diquarks 
dominate the beam fragmentation region. 
The most well known process of this type is 
$p+p\to \Lambda +X$. 
To see whether, if yes how, we can also understand the existence 
of $P_\Lambda$ in this process, 
we start again from the single-spin process 
$p(\uparrow)+p\to \Lambda+X$. 
We recall that the recent E704 data [5,6] shows that, 
also for $\Lambda$, 
there exists a significant $A_N$ 
in the beam fragmentation region.  
At the first sight, 
this result seems rather surprising because  $\Lambda$ 
in the beam fragmentation region comes predominately from 
the hadronization of the spin-0 $(u_vd_v)^P$ 
of the projectile. 
How can a spin-0 object transfer 
the information of polarization of the projectile  
to the produced $\Lambda$?
This question has been discussed [12] 
and a solution has been suggested in which  
associated production plays an important role. 
It has been pointed out [12] that, 
the production of the $\Lambda$ containing 
the spin-0 $(u_vd_v)^P$ and a $s_s^T$ is
associated with the production of a Kaon 
containing the remaining 
$(u_v^a)^P$ of the projectile 
and the $\bar s^T_s$ associated with the target.  
The information of polarization of the projectile
is carried by the $(u_v^a)^P$
so that the produced $K$ has a large probability to 
go left if the projectile is upwards polarized. 
The $\Lambda$ has therefore a large probability to go
right since the transverse momentum should be compensated. 
This explains why there should be also 
a significant $A_N$ for $\Lambda$, 
and the available data [5,6] has been reproduced successfully.
According to this picture,  
if the produced $\Lambda$ is moving to the
left in unpolarized $pp$-collision, 
the associated $K$ should mainly move to the right. 
Hence, the $(u_v^a)^P$ contained in this $K$ should have 
a large probability to be downwards polarized. 
Since $K$ is a spin-0 object, the $\bar s_s^T $ should
be upwards polarized. 
Hence, to get a negative $P_\Lambda$, 
we need only to assume 
that the sea quark-antiquark pair $s_s\bar s_s$ 
from the nucleon have opposite transverse spins.  
Under this assumption, the polarization of the produced 
$\Lambda $ is completely determined [17] by that 
of the remaining $(u_v^a)^P$ which, 
together with a $\bar s_s^T$, forms the associatively 
produced $K^+$.  

That the sea-quark-antiquark pair $s_s\bar s_s$ 
have opposite transverse spins should be considered as 
a further implication of the existence of $P_\Lambda$ 
in the above mentioned picture. 
Whether this is indeed the case can and 
should be checked experimentally. 
Theoretically, 
it is quite difficult to verify it
since we are in the very small $x$ region (see e.g. [12]), 
the production of such pairs 
is of soft nature in general and  
cannot be calculated 
using perturbative theory.
It seems plausible since  
the sea quarks are products of the dissociation 
of one or more gluons and gluons 
are not transversely polarized.
Here, we simply assume this is true and 
discuss the consequences to see whether they are 
consistent with the available data.

First, we made a
similar analysis for the production of other hyperons, and  
obtained qualitative results for their $P_H$'s.   
They are all consistent with the available data~[2]. 

Second, we made a quantitative estimation 
of $P_\Lambda$ in $p+p\to\Lambda+X$ 
as a function of $x_F$. 
To do this, we recall that $P_\Lambda (x_F|s)$ is 
defined as 
\begin{equation}
P_\Lambda(x_F|s)\equiv 
 {N^\Lambda(x_F,\uparrow |s)-N^\Lambda(x_F,\downarrow |s)\over 
  N^\Lambda(x_F,\uparrow |s)+N^\Lambda(x_F,\downarrow |s)},
\end{equation}
where $N^\Lambda(x_F,i |s)$ 
is the number density of $\Lambda$'s 
polarized in the same $(i=\uparrow )$ 
or opposite $(i=\downarrow )$  direction as the 
normal of the production plane, 
at a given $\sqrt{s}$.
It is clear that the denominator is nothing else but the 
number density of $\Lambda$
without specifying the polarization. 
It contains all the $\Lambda$'s 
of different origins: 
those made out of $(u_vd_v)^P$ and a $s_s^T$ 
[denoted by $D_2^\Lambda(x_F|s)$ in the following]; 
those of $u_v^P$ and $(d_ss_s)^T$ or $d_v^P$ and $(u_ss_s)^T$ 
[denoted by $D_1^\Lambda(x_F|s)$];   
those from resonances decay; 
and those from pure sea-sea interactions, 
[denoted by $N_0$].
Since $u_v$ or $d_v$ does not carry any 
information of the spin of $\Lambda$, 
there is no contribution from $D_1^\Lambda(x_F|s)$ to  
the numerator, i.e., the difference 
$\Delta N^\Lambda(x_F|s)$.
There is no contribution to $A_N$ from the $N_0$ part, 
hence we assume that it does not contribute 
to $P_H$ either [18]. 
We take the contribution from 
$\Sigma^0$-decay 
into account and obtain, 
\begin{equation}
\Delta N^\Lambda(x_F|s)= C [\Delta D_2^\Lambda(x_F|s)
       -{1\over 3} \sum_{i=1}^{2}
       \Delta D_i^{\Sigma^0}(x_F|s)]. 
\end{equation}
Here $\Delta D_i^H(x_F|s)\equiv 
D_i^H(x_F,{\uparrow} |s) -D_i^H(x_F,{\downarrow} |s)$, 
($H=\Lambda$ or $\Sigma^0$). 
From the wave functions of 
$\Lambda$, $\Sigma^0$ and that of proton, 
we obtain that
$\Delta D_2^\Lambda(x_F|s)=-D_2^\Lambda(x_F|s)$;
and $\Delta D_{1,2}^{\Sigma^0}(x_F|s)=
(\frac{2}{3},\frac{3}{5})D_{1,2}^{\Sigma^0}(x_F|s)$. 
The extra factor $-1/3$ for the $\Sigma^0$-decay terms 
comes from the relation [19] 
$P_\Lambda =-(1/3)P_{\Sigma^0}$ 
in this decay process. 
To calculate the different $D$'s and $N_0$, 
which are determined by the hadronization mechanisms,   
we simply used the direct fusion model in [12], 
which successfully reproduced 
not only the data of the cross section but also those of $A_N$.
By taking the same value for  the only free parameter $C$,   
as that determined in [11,12] by fitting the $A_N$ data [5,6], 
we obtained the result shown in Fig.1.

Third, we derived a number of 
other consequences 
of the picture without any further input. 
The following are three examples which are 
closely related to the assumption that the 
$s$ and $\bar s$ which take part in the associated production 
are opposite in transverse spins.

($\alpha$) The polarization of the projectile 
and that of $\Lambda$ in the fragmentation region 
of $p+p\to \Lambda +X$ should be closely 
related to each other. 
In other words, the spin transfer 
$D_{NN}$ (i.e. the probability 
for the produced $\Lambda$ to be upwards polarized 
in the case that the projectile proton is upwards polarized)
is expected to be positive 
and large for large $x_F$.   
It is true that the $ud$-diquark 
which forms the $\Lambda$ is in 
a spin-zero state thus carries no 
information of polarization.  
But, according to the mechanism of 
associated production, 
the polarization of the left-over $u_v^P$ 
determines the polarization of the projectile and that 
of the $s_s$ quark which combines with the 
$ud$-diquark to form the $\Lambda$. 
Hence, there should be 
a strong correlation between the polarization of 
the proton and that of the $\Lambda$. 
The result of a quantitative estimation  is shown in Fig.2.

($\beta$) $P_\Lambda$ in the beam fragmentation region 
of $\Sigma^-+A\to \Lambda +X$ 
should be {\it negative} and much less significant 
than that in $p+p\to\Lambda+X$.  
Here, the dominating contributions are 
the $\Lambda$'s which consist of 
$(d_vs_v)^P$ and $u_s^T$, 
$d_v^P$ and $(u_ss_s)^T$,  
or $s_v^P$ and $(u_sd_s)^T$.
Exactly the same analysis as  
above for $p+p\to\Lambda+X$ shows that
the $\Lambda$'s of the first two kinds are unpolarized; 
and those of the third kind   
are positively polarized. 
Hence, if we exclude the contribution from 
$\Sigma^0$ and $\Sigma^{*0}$ decay, 
$P_\Lambda$ should be approximately zero for 
large $x_F$ and should be small but 
positive in the middle $x_F$ region. 
Taking $\Sigma^0$ and $\Sigma^*$ decay into account, 
we expect a small negative $P_\Lambda$ for large $x_F$. 

($\gamma$) Hyperon polarization in  
processes in which a vector meson is associatively produced 
should be very much different from that in  
processes in which a pseudoscalar 
meson is associatively produced.  
E.g., $P_\Lambda$ in the fragmentation region of   
$p+p\to \Lambda +K^{+}+X$ should be negative 
and its magnitude should be large, but  
$P_\Lambda$ in the fragmentation region of  
$p+p\to \Lambda +K^{*+}+X$ should be positive 
and its magnitude should be much smaller.
This is because, in the latter case, 
using the same arguments as we used in the former case,
we still obtain that $(u_v^a)^P$ 
(contained in $K^{*+}$) 
has a large probability to be downwards polarized 
if $\Lambda $ is going left.  
But, in contrast to the former case, 
the $\bar s_s^T$ here in the $K^{*+}$ can be upwards 
or downwards polarized since $K^{*+}$
is a spin-1 object. 
If $\bar s_s^T$ is upwards polarized, 
the produced meson can either be a $K^*$ or a $K$, 
and the corresponding $\Lambda$ should be 
downwards polarized. 
But if $\bar s_s^T$ is downwards polarized, 
the produced meson can only be a $K^{*+}$ and 
the corresponding $\Lambda$ should 
be upwards polarized, i.e. $P_\Lambda>0$.

Presently, there are data available 
for the processes mentioned 
in ($\alpha$) and ($\beta$) [6,23], 
and both of them are in agreement with 
the above expectations. 
The prediction mentioned in ($\gamma$) 
is another characteristic feature of the model 
and can be used as a crisp test of the picture.

We thank K. Heller, Meng Ta-chung and 
R. Rittel for helpful discussions.
This work was supported in part by Deutsche
Forschungsgemeinschaft (DFG:Me 470/7-2).

\newpage


\begin{figure}
\psfig{file=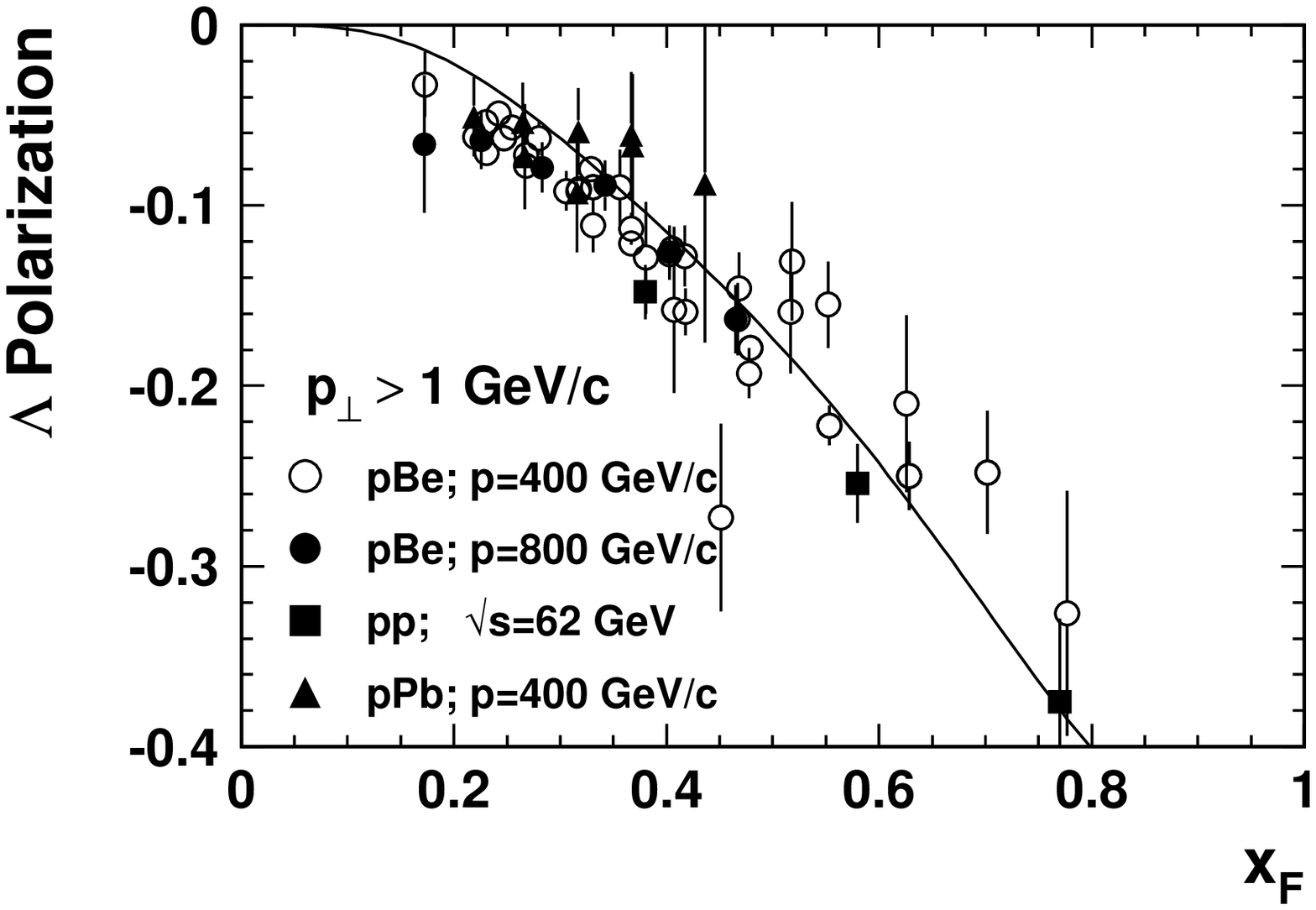,height=6cm}
\caption{Calculated results of the polarization of $\Lambda$,
$P_\Lambda$, as a function of $x_F$.
Data are taken from [20-22].}
\end{figure}

\begin{figure}
\psfig{file=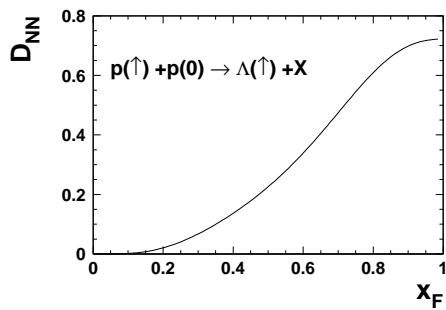,height=5cm}
\caption{$D_{NN}$ as a function of $x_F$
calculated using the proposed picture
for the case that the correlation between the
spin of the $s_s^T$ [which forms together with the
$(u_vd_v)_{0,0}^P$ the $\Lambda$]
and the spin of the remaining $u_v^P$ of the projectile
(which forms together with the $\bar s_s^T$
the associated $K^+$) is maximal.
In this sense, it stands for the
upper limit of our expectation. }
\end{figure}


\begin{references}
\bibitem{foot} We note that this correlation remains unity even 
               if more pseudoscalar mesons are created, 
               but may be destroyed if vector mesons 
               are associatively produced. 
               Here, we consider only the former case. 
               In this sense, what we obtain is the upper limit 
               of the expectation from the picture. 
               See in this connection also the point 
               $(\gamma)$ later in this paper.
\bibitem{foot} This does not necessary mean that the sea quarks 
              are absolutely unpolarized. 
              It may just mean that their relative polarization 
              is negligibly small or they lose their information 
              of polarization during the initial state interactions 
              between the constituents of the two colliding hadrons.
\end{references}
\end{document}